
\input phyzzx
{\hsize=17.5truecm \leftskip=11.5cm
{INS-Rep.-903}\par
\vskip -4mm
{OCHA-PP-23}\par
\vskip -4mm
{December 1991}\par}
\def\pref#1{\rlap.\attach{#1}}
\def\cref#1{\rlap,\attach{#1}}
\def\odd{{\rm odd}}
\def\ie{{\it i.e.}}

\def\LA{\ \Longrightarrow\ }
\def\e{{\rm e}}
\def\gz{\gamma_0}
\def\lz{\lambda\gamma_0^2}
\def\tr{{\rm tr}}
\def\co{{\cal O}}
\def\ln{{\rm ln\,}}

\title{Non-perturbative Aspect of Zero Dimensional Supersring}
\author{Shin'ichi Nojiri\foot{INS fellow}}
\address{Institute for Nuclear Study, University of Tokyo}
\address{Tanashi, Tokyo 188, Japan}
\centerline{and}
\address{Faculty of Science, Department of Physics,
Ochanomizu University}
\address{1-1 Otsuka 2, Bunkyo-ku, Tokyo 112, JAPAN}

\abstract{We discuss the non-perturbative aspect of zero dimensional
superstring. The perturbative expansions of correlation functions
diverge as $\sum_l(3l)!\kappa^{2l}$, where $\kappa$ is a string coupling
constant. This implies there are non-perturbative contributions of order
$\e^{C\kappa^{-{2 \over 3}}}$. (Here $C$ is a constant.) This situation
contrasts with those of critical or non-critical bosonic strings,
where the perturbative expansions diverge as $\sum_ll!\kappa^{2l}$ and
non-perturbative behaviors go as $\e^{C\kappa^{-1}}$.
It is explained how such nonperturbative effects of order
$\e^{C\kappa^{-{2 \over 3}}}$ appear in zero dimensional superstring
theory. Due to these non-perturbative
effects, the supersymmetry in target space breaks down spontaneously.}

\endpage
\sequentialequations

\REF\ri{D.J. Gross and V. Periwal, \journal Phys.Rev.Lett. &60 (88)
2105, \journal Phys.Rev.Lett. &61 (88) 1517}
\REF\rii{M.R. Douglas and S.H. Shenker, \journal Nucl.Phys. &B335 (90) 635}
\REF\riii{D.J. Gross and A.A. Migdal, \journal Phys.Rev.Lett. &64 (90) 127}
\REF\riv{E. Br\'ezin and V.A. Kazakov, \journal Phys.Lett. &B236 (90) 144}
\REF\rv{E. Marinari and G. Parisi, \journal Phys.Lett. &B240 (90) 375}
\REF\rvii{S. Nojiri, \journal Phys.Lett. &B253 (91) 63 \journal
Prog.Theor.Phys. &85 (91) 671}
\REF\rvi{S. Nojiri, \journal Phys.Lett. &B252 (90) 561 }
\REF\rvia{A. Dabholkar, Rutgers preprint RU-91/20 (91)}
\REF\rvib{H. Nicolai, \journal Phys.Lett. &B101 (80) 341 \journal
Nucl.Phys. &B176 (80) 419}
\REF\rxvi{V. Kazakov, \journal Phys.Lett. &B150 (85) 282}
\REF\rxvii{J. Ambj\o rn, B. Durhuus and J. Fr\"ohlich, \journal
Nucl.Phys. &B257 (85) 433}
\REF\rxviii{V. Kazakov, I. Kostov and
A. Migdal, \journal Phys.Lett. &B157 (85) 295}
\REF\rrrii{V. Kazakov, I. Kostov and
A. Migdal, \journal Nucl.Phys. &B275 (86) 641}
\REF\rrriii{V. Boulatov, V. Kazakov, I. Kostov
and A.A. Migdal, \journal Nucl.Phys. &B275 (86) 641}
\REF\rrriv{F. David, \journal Phys.Lett. &B159 (85) 303}
\REF\rrrv{I. Kostov and M.L. Mehta, \journal Phys.Lett. &B189 (87) 118}
\REF\rrrvi{G. Parisi and N. Sourlas, \journal Phys.Rev.Lett. &43 (76) 74}


One of most important problems in superstring theories is how their
supersymmetries break down. It seems that the spontaneous breakdown of
supersymmetry never happens at classical level or perturbative level.
In case of bosonic string, Gross and Periwal have proved that
the perturbation theory diverges and is not Borel summable\pref\ri
It points out that the perturbative vacuum is unstable and the true
vacuum is picked out by non-perturbative dynamics. This suggests that
non-perturbative analysis will be neccesary in order to solve the
problem of the spontaneous breakdown of the supersymmetry in
superstring theories. Few years ago, a great progress was made in the
non-perturbative formulation of string theories in less than one
dimension\pref{\rii - \riv}
These models can be regarded as an important solvable \lq toy' model,
which may give a clue to solve the dynamics of \lq realistic' string
models. In this development, string models which have a supersymmetry
in one dimension was proposed by Marinari and Parisi\pref\rv
After that, the author constructed superstring models in less than
one dimension\refmark\rvii and zero dimension\pref\rvi In this paper,
we analyze zero dimensional superstring, which is the simplest
superstring model. The possibility of the spontaneous breakdown
of the supersymmetry has already been suggested in Ref.\rvi .
We now discuss the spontaneous breakdown of the supersymmmetry by
investigating the free energy. The perturbative expansions of
correlation functions in zero dimensional superstring theory diverge
as $\sum_l(3l)!\kappa^{2l}$. Here $\kappa$ is a string coupling
constant. This implies that there will appear the non-perturbative
contributions of order $\e^{C\kappa^{-{2 \over 3}}}$.
Here $C$ is a constant. This forms a strange contrast to the cases of
critical or non-critical bosonic string theories. The perturbative
expansion of the free energy $F$ in critical
bosonic string is given by $F\sim\sum_ll!\kappa^{2l}$ and we expect
the non-perturbative effects of order $\e^{C\kappa^{-1}}$\cref\ri which
have been also observed in case of non-critical bosonic strings\pref\riii
Similar non-perturbative effects were also found in Marinari-Parisi's
one dimensional superstring theory\pref\rvia
In this paper, it will be explained how the non-perturbative effects of
order $\e^{C\kappa^{-{2 \over 3}}}$ appear in zero dimensional
superstring theory and how the supersymmetry breaks down spontaneously.

The partition function of zero dimensional superstring theory is given
by the path integrals of an $N \times N$ hermitian matrix $A_{ij}$
$(i,j=1,\cdots, N)$ and fermionic (anti-commuting) $N \times N$
hermitian matrices $\overline \Psi_{ij}$, $\Psi_{ij}$
$(i,j=1,\cdots, N)$.
$$Z=(\lambda N)^{-N^2}\int dAd\overline \Psi d\Psi \exp
\lambda S(A,\overline \Psi , \Psi) \ .\eqn\i$$
The action $S(A,\overline \Psi , \Psi)$ has the following form:
$$S(A,\overline \Psi , \Psi)=N \lbrace
-{1 \over 4}{\rm tr} ({\partial W(A) \over \partial A})^2
- {1 \over 2}\sum_{i,j,k,l=1}^N\overline \Psi_{ij} \Psi_{kl}
{\partial^2 W(A) \over \partial A_{ij}
\partial A_{kl}} \rbrace \ .\eqn\ii$$
Here $W(A)$ is a superpotential,
$$W(A)=\sum_{l=1}^Lg_l {\rm tr} A^l \ .\eqn\iii$$
The factor $(\lambda N)^{-N^2}$ in Eq.\i\ appears due to the
integration of the auxilliary field. The system is invariant under the
following supersymmetry transformation in zero dimension:
$$\eqalign{\delta A=&\overline \epsilon \Psi
+ \epsilon \overline \Psi \ , \cr
\delta \Psi=&
{1 \over 2}\epsilon {\partial W \over \partial A} \ , \cr
\delta \overline \Psi=&
-{1 \over 2}\overline \epsilon
{\partial W \over \partial A} \ .}\eqn\iv$$
By using the Nicolai mapping\refmark\rvib
$$\Gamma={1 \over 2}
{\partial W(A) \over \partial A} \ ,\eqn\v$$
it has been shown that the invariance under the transformation \iv\
guarantees that the free energy $F=\ln Z$ of the matrix model, {\it i.e.},
the vacuum amplitude of the corresponding string theory, vanishes in
any order of the perturbation, $F=0$. In case of $L=\odd$, however,
the supersymmetry breaks down spontaneously, and the non-perturbative
partition function vanishes, $Z=0$, and the free energy goes to
infinity, $F\rightarrow \infty$.

In Ref.\rvi , we have found that there exists a critical point by
analyzing the correlation functions $<{\rm tr} A^m >$. The critical
point appears in the large $N$ limit when the Nicolai mapping in Eq.\v\
is degenerate:
$$\Gamma =-{1 \over n g^{n-1}}(g-A)^n+{g \over n} \ .\eqn\vi$$
Here $g$ is a coupling constant. If we define $x$ by
$$x={2^2 n^2 \over g^2 \lambda }\ ,\eqn\via$$
the correlation functions $<{\rm tr} A^m >$ has the following form
when the Nicolai mapping is degenerate \vi ,
$${1 \over g^m N}<{\rm tr} (g-A)^m >
\sim \sum_l c_l N^{-2l}(1-x)^{-3l+{3 \over 2}+{m \over n}}\ . \eqn\vii$$
Therefore, if we fix
$$\kappa^{-1}= N(1-x)^{3/2} \eqn\viii$$
then by letting $x \rightarrow 1$ as $N \rightarrow \infty$, we obtain
finite correlation functions up to multiplicative renormalization
constants to all orders in the ${1 \over N}$ ({\it i.e.} genus)
expansion and $\kappa$ can be regarded as a renormalized string
coupling constant. An interesting point is the behavior of the
coefficients $c_l$ in Eq.\vii . By using the formulae in Ref.\rvi ,
we can easily find, when $l$ is large,
$$c_l\sim (3l)!\ . \eqn\ix$$
The perturbative expansions of the correlation functions diverge as
$\sum_l(3l)!\kappa^2$. Therefore we expect that there will appear
non-perturbative contributions of order $\e^{C\kappa^{-{2 \over 3}}}$.
In the following, we consider how such contributions appear in case
$L$ in Eq.\iii\ is odd ($n$ in Eq.\vi\ is even) \ie, in case the
supersymmetry breaks down spontaneously.

When $L=\odd$, the partition function $Z$ vanishes, therefore the
expectation value of any operator diverges or vanishes in general and
the non-perturbative theory is ill-defined. In order to obtain a
well-defined theory, we modify the Nicolai mapping in Eq.\vi\ as
follows:
$${\partial W(A) \over \partial A}=\Gamma
=-{(\e^{mg}-\e^{mA})^n \over nm(\e^{mg}-1)^{n-1}}
+{\e^{mg}-1 \over nm}\ .\eqn\x$$
This modification does not change the behavior when $A\sim g$\foot{
To be exact, when an eigenvalue $a$ of the matrix $A$ approaches to $g$,
the behavior of the Nicolai mapping does not change in the leading order
{\it w.r.t.} $a-g$ under the modification. The behavior is only relevant
to the critical behavior of the correlation functions.}
and the modified Nicolai mapping \x\ reduces to the previous one in the
limit of $m\rightarrow 0$. When $A$ goes to $+\infty$, $\Gamma $ goes
to $-\infty$ but when $A$ goes to $-\infty$, $\Gamma $ remains to be
finite:
$$\eqalign{A\rightarrow -\infty\ \LA\ &\Gamma\rightarrow -\gamma_0\cr
&\gamma_0\equiv{\e^{nmg} \over nm(\e^{mg}-1)}-{\e^{mg}-1 \over nm}}
\eqn\xa$$
Therefore the partition function of the modified theory is finite when
$m$ is finite.

The modified theory can be regarded as a regularized theory of the
original theory \vi . The modification \x\ does not change the
critical properties of the original theory. If we define
$$\tilde A\equiv {\e^{mA}-1 \over m}\ ,\ \
\tilde g\equiv {\e^{mg}-1 \over m}\ , \eqn\xi$$
the Nicolai mapping \x\ can be rewritten by
$$\Gamma =-{1 \over n \tilde g^{n-1}}
(\tilde g-\tilde A)^n+{\tilde g \over n}\ .\eqn\xii$$
Then by using the argument given in Ref.\rvi , we can find that the
correlation function
${1 \over \tilde g^m N}<{\rm tr} (\tilde g-\tilde A)^m >$ shows
the same critical behavior as the previous one in Eq.\viii :
$${1 \over \tilde g^m N}<{\rm tr} (\tilde g-\tilde A)^m >
\sim \sum_l c_l N^{-2l}(1-x)^{-3l+{3 \over 2}+{m \over n}}\ .
\eqn\xiib$$
Here we have redefined $x$ by
$$x\equiv{2^2n^2m^2 \over (\e^{mg}-1)^2\lambda}\ . \eqn\xiia$$
Therefore the modified theory belongs to the same universality class as
the original one.

Since the modification \x\ is given in terms of the Nicolai mapping,
\ie, the derivative of superpotential $W(A)$ with respect to $A$,
the perturbative supersymmetry corresponding to Eq.\iv\ remains in the
modified theory. The supersymmetry, however, breaks down spontaneously
since the finite modified partition function depends on the coupling
constants.
In the following, we invetigate if the breakdown of the supersymmetry
remains after the double scaling limit and how the non-perturbative
contribution of order $\e^{C\kappa^{-{2 \over 3}}}$ appears.

The modified partition function is given by
$$\eqalign{Z&=( 2\lambda N)^{N^2 \over 2}
\int d\Gamma \exp \lbrace - \lambda N {\rm tr} \Gamma
\Gamma \rbrace \cr
&=( 2\lambda N)^{N^2 \over 2}
\int_{\gamma_0}^\infty \prod_{n=1}^N d\gamma_n
\prod_{m>l} (\gamma_m-\gamma_l)^2
\exp (-{1 \over 2}\sum_{k=1}^N \gamma_k^2)}\eqn\xiii$$
Here we have diagonalized the matrix $\Gamma $ by the unitary matrix $U$,
$$\Gamma ={1 \over \sqrt{ 2\lambda N}}U^{-1}\gamma U\ ,\ \
\gamma={\rm diag}(\gamma_1, \gamma_2, \cdots , \gamma_N)\ .\eqn\xiv$$
By changing the variable : $\gamma_i={1 \over 2\lambda N\gz}y_i+\gz$,
the Equation \xiii\ can be rewritten by
$$\eqalign{Z=&N^{-{N^2 \over 2}}(\lz)^{-{N^2\over 2}}\e^{-N^2\lz}\cr
&\times \int_0^\infty \prod_{n=1}^N dy_n
\prod_{m>l} (y_m-y_l)^2
\exp \{-\sum_{k=1}^N (y_k+{y_k^2 \over 4N\lz})\}\cr
=&N^{-{N^2 \over 2}}(\lz)^{-{N^2\over 2}}\e^{-N^2\lz}
\int [d\Phi]\, \exp \{-\tr\Phi-{\tr\Phi^2 \over 4N\lz}\}\cr
=&N^{-{N^2 \over 2}}(\lz)^{-{N^2\over 2}}\e^{-N^2\lz}
Z_0<\exp \{-{\tr\Phi^2 \over 4N\lz}\}>_0}\eqn\xv$$
Here $\Phi$ is a hermitian matrix whose eigenvalues are positive
semi-definite. We define $Z_0$ and $<\cdots>_0$ by
$$Z_0\equiv\int [d\Phi]\, \exp \{-\tr\Phi\},\
<{\cal O}>_0\equiv Z_0^{-1}\int [d\Phi]\, {\cal O}\,\exp \{-\tr\Phi\}\ .
\eqn\xvi$$
Since the orthogonal polynomials whose measure is given by
$\int_0^\infty dx\e^{-x}\cdots$ are Laguerre's polynomials,
$$L_n(x)\equiv \sum_{m=0}^n(-1)^m{n \choose n-m}
{x^m \over m!}\ ,\eqn\xvia$$
we can calculate $Z_0$ straightforwardly,
$$Z_0=N!\prod_{n=1}^{N-1}(n!)^2\ .\eqn\xvii$$
Furtheremore we know the general properties of expectation values
$$\eqalign{<\e^\co>&\leq\e^{<\co>}\ \ \
\ \ ({\rm convex\ inequality})\cr
<\e^\co>&=\e^{<\co>+{1 \over 2!}<\co^2>_c
+{1 \over 3!}<\co^3>_c+\cdots}\cr
&<\co^2>_c\equiv <\co^2>-<\co>^2\cr
&<\co^3>_c\equiv <\co^3>-3<\co><\co^2>_c-<\co>^3\cr
&\ \ \cdots
\ .}\eqn\xviii$$
Therefore if we define
$$z\equiv\lz\ ,\eqn\xviiia$$
we find
$$<\e^{-{1 \over 4Nz}\tr\Phi^2}>_0
=\e^{-{1 \over 4Nz}<\tr\Phi^2>_0
+{1 \over 2!}({1 \over 4Nz})^2<\tr\Phi^2\tr\Phi^2>_c
+\co(z^{-3})}\ .\eqn\xix$$
Explicit calculation gives
$$<\tr\Phi^2>_0=2N^3
\ \  ({\rm no\ higher\ order\ terms}\ w.r.t.\ {1 \over N})\ .\eqn\xx$$
We can also estimate $<(\tr\Phi^2)^n>_c$ by using the factorization
properties,
$$<(\tr\Phi^2)^n>_c\sim\co(N^{3n-2(n-1)})\sim\co(N^{n+2})\ .\eqn\xxi$$
Then the partition function is given by
$$\eqalign{Z&=N^{-{n^2 \over 2}}N!\prod_{n=1}^{N-1}(n!)^2
\e^{-N^2f(z)-g(z)+\co(N^{-2})}\cr
f(z)&=z+{1 \over 2}\ln z+{1 \over 2z}+\co(z^{-2})\cr
g(z)&=\co(z^{-2})\ .}\eqn\xxii$$
Due to convex inequality, the function $f(z)$ is bounded below by
$$f(z)\geq z+{1 \over 2}\ln z+{1 \over 2z}\eqn\xxiii$$
In the following, we consider how the non-perturbative effect
of order $\e^{C\kappa^{-{2 \over 3}}}$ appears.

We begin with counting how many parameters (coupling constants) this
theory has. We have three parameters $(\lambda, g, m)$ at first but one
of these parameters are redundant since we can redefine (or rescale)
the matrix field $A$ by $A\rightarrow \e^t A$. ($t$ is a parameter of
rescaling. Two parameters which are invariant under this redefinition
are given by $(x, z)$ in Eqs.\xiia\ and \xviiia . Furtheremore
we know that the universality class does not change if we vary the
parameter $m$. Therefore there will be one more redundant parameter
which we denote by $\Lambda $. Since $x$ is apparently a parameter
specifying the theory, $z$ is a function of $x$ and $\Lambda $ in
general:
$$z=z(x,\Lambda )\ .\eqn\xxiv$$
Since $x=1-(N\kappa)^{-{2 \over 3}}$ (Eq.\viii ), we can expand
$N^2f(z)$ in Eq.\xxii\  when $N$ is large:
$$N^2f(z(x,\Lambda ))=N^2 f(z(1,\Lambda))+N^{4 \over 3}\partial_zf(z(1,
\lambda))\partial_xz(1,\Lambda )\kappa^{-{2 \over 3}}
+\cdots\eqn\xxv$$
The first term $N^2 f(z(1,\Lambda))$ is essentially c-number since this
term only depends on the redundant parameter $\Lambda $. Therefore this
term can be absorbed into the renormalization of the matrix field $A$
and does not contribute to the expectation value of any operator. The
second term $N^{4 \over 3}\partial_zf(z(1,\lambda))
\partial_xz(1,\Lambda )\kappa^{-{2 \over 3}}$, however, has a physical
meaning. We can make this term finite by adjusting the redundant
parameter $\Lambda $. Therefore we can find that a non-perturbative
contribution of order $\e^{\kappa^{-{2 \over 3}}}$ appears in the
partition function,
$$Z\sim\e^{C\kappa^{-{2 \over 3}}}\ .\eqn\xxvi$$
There is an ambiguity or freedom how to choose the redundant parameter
$\Lambda $. For example, we can choose $\Lambda $ by
$$z=c_1-c_2{x \over \Lambda}\ ,\eqn\xxvii$$
here $c_1$ and $c_2$ are arbitrary constants. If we adjust
$\Lambda\sim N^{2 \over 3}$, we obtain a finite partition function
and a finite free energy. Since the free energy does not vanish,
the supersymmetry in zero dimensional target space breaks down spontaneously.

In summary, we have investigated the non-perturbative aspect of zero
dimensional superstring. For this purpose, we have proposed a kind of
regularization which is given inEq.\x . This regularization has
following properties.
\item{1)} This regularization makes the partition function to be finite
when $N$ is finite.
\item{2)} The regularized model belongs to the same universality class
as the original model.
\item{3)} This regularization keeps the supersymmetry perturbatively.
\item{3)} This regularization breaks the supersymmetry
non-perturbatively.

\noindent
By using this regularization, it has been explained how the
non-perturbative contributions of order $\e^{C\kappa^{-{2 \over 3}}}$
appear. This contribution is consistent with the perturbative
expansions of the correlation functions $\sim\sum_l(3l)!\kappa^{2l}$.
Due to these contributions, the free energy does not vanish and the
supersymmetry in zero dimensional target space breaks down spontaneously.
These results do not depend on the details of the regularization.
Any regularization which has the properties 1)--2) apparantly gives
the same results.

The zero dimensional superstring theory is closely related to the bosonic
two dimensional gravity coupled with $c=-2$ conformal matter
($-2$ dimensional string theory)\pref{\rxviii - \rrrv}
The $-2$ dimensinal string theory can be obtained from zero dimensional
string theory by Parisi and Sourlas's dimensional reduction
mechanism\pref\rrrvi
Parisi and Sourlas's mechanism connects $D$ dimensinal theory to $D-2$
dimensional one. The correlation functions of $D$ dimensinal theory,
except vacuum amplitude, are identical with those of
$D-2$ dimensional one if the support of $D$ dimensional
correlation functions is restricted to
$D-2$ hypersurface. Of course, this does not mean that
these two theories are equivalent. We expect, however, the
non-pertubative effects observed in this paper will also appear in the
bosonic strings in $-2$ dimensions.

I would like to acknowledge M. Kato and A. Sugamoto for discussions.

\refout
\bye